# Encapsulated silicon nitride nanobeam cavity for nanophotonics using layered materials

Taylor K. Fryett[1], Yueyang Chen[1], James Whitehead[1], Zane Matthew Peycke[2], Xiaodong Xu[2,3], Arka Majumdar[1,2]

[1] Electrical Engineering, University of Washington, Seattle, WA 98189, USA

[2] Department of Physics, University of Washington, Seattle, WA 98189, USA

[3] Materials Science and Engineering, University of Washington, Seattle, WA 98189, USA

ABSTRACT:

Most existing implementations of silicon nitride photonic crystal cavities rely on suspended membranes due to the low refractive index of silicon nitride. Such floating membranes are not mechanically robust, making them suboptimal for developing a hybrid optoelectronic platform where new materials, such as layered 2D materials, are transferred on a pre-existing optical cavity. To address this issue, we propose a silicon nitride nanobeam resonator design where the silicon nitride membrane is encapsulated by material with a refractive index of ~1.5, such as silicon dioxide or PMMA. The theoretically calculated quality factor of the cavities can be as large as $10^5$, with a mode-volume of ~$2.5\left(\frac{\lambda}{n}\right)^3$. We fabricated the cavity, and measured the transmission spectrum with highest quality factor of 7000. We also successfully transferred monolayer tungsten diselenide on the encapsulated silicon nitride nanobeam, and demonstrated coupling of the cavity with the monolayer exciton and the defect emissions.

TEXT

**Introduction**

Silicon nitride (SiN) offers several advantages over silicon for building photonic integrated circuits due to its large band-gap. For instance, two-photon absorption in SiN is negligible at the telecommunication wavelengths. This allows operation at much higher optical power, and with significantly lower loss compared to similar silicon devices. The thermo-optic effect in SiN is also an order of magnitude smaller compared to silicon, and could potentially provide a scalable integrated photonic platform that is far less susceptible to thermal fluctuations. However, a major problem with SiN is that its carrier densities cannot be modulated easily due to its large bandgap, resulting in a lack of active devices. This problem is worsened by the amorphous nature of SiN grown via plasma enhanced, and low pressure chemical vapor deposition (PECVD, and LPCVD) which makes integration of other active materials by epitaxial growth more difficult. For example, complex electro-optic oxides[1] or quantum confined structures[2], which can be grown or wafer-bonded on silicon, cannot be integrated on SiN without compromising the performance. In contrast, layered 2D materials including graphene and transition metal dichalcogenides (TMDCs) can adhere to any substrate via van der Waals forces. This removes the explicit lattice matching requirement of epitaxial growth[3, 4].

Recently demonstrated micro-ring electro-optic modulators have shown that layered materials provide a promising method to build hybrid active photonic systems in SiN[5]. Transition metal dichalcogenides have also been integrated on SiN micro-ring and disk resonators to explore the Purcell effect[6-8]. However, all previous work on 2D material clad SiN resonators has focused on whispering gallery resonators. Photonic crystal resonators are often preferred to enhance light-matter interaction due to their small mode-volumes ($V_m$). Photonic crystal resonators made of silicon and gallium phosphide have already been used to demonstrate cavity enhanced photoluminescence (PL)[9-11], electroluminescence[12], and second harmonic generation[13, 14] in layered materials. Unfortunately, the small refractive index of SiN (n~2) inhibits the opening of a complete band gap in two-dimensional photonic crystals. This is particularly true for common crystal lattices such as hexagonal and square lattices. This difficulty is the primary reason why many in the community use nanobeam resonators, floating 1D photonic crystal structures where bandgaps are more readily opened[15].

An outstanding challenge presented by the floating membranes is its propensity to be damaged by common micro-fabrication techniques such as resist spinning for lithographic overlay. The popular dry transfer techniques for building van der Waals heterostructures[16, 17] can also easily destroy the floating membranes (see Supplementary Materials). While, one can realize hybrid devices without transferring the layered materials, as recently reported[16], it is difficult to integrate different 2D materials on the same photonic chip and the resulting floating membranes still remain susceptible to damage from further fabrication steps. Finally, for rapid prototyping and initial characterization of the layered material clad cavities, the 2D material must be removed by vigorous sonication. Therefore, floating membranes are not optimal for use in a hybrid photonic platform.

In this Letter, we design and fabricate a SiN nanobeam cavity encapsulated inside a medium of refractive index 1.5. Such an encapsulation with different refractive index material is crucial for practical operation as well as desirable for long-term stability of layered materials, including black phosphorous[17] and MoTe$_2$[18] materials with light-emission in the infrared regime. Via numerical simulations, we found that the cavity by our design can reach a quality (Q) factor of ~$10^5$ and a mode volume of $\sim 2.5 \left(\frac{\lambda}{n}\right)^3$. Experimentally, we measured a Q-factor of up to $\sim 7,000$ in transmission. We further demonstrated integration of monolayer tungsten diselenide (WSe$_2$) with the nanobeam cavity and observed the cavity enhanced PL. Our work thus presents a new way towards hybrid 2D material – cavity photonic circuits using low mode-volume SiN resonators.

**Design of the resonator**

To design the nanobeam optical resonators, we first simulated the band structure of a SiN ($n \sim 2$) 1D photonic crystal with surrounding medium of $n \sim 1.5$ using the MIT photonic bands (MPB) software package[19]. We assume that the holes are elliptical following a previously reported design[20]. Figure 1a shows the resulting band diagram which includes the air-band, the dielectric band, and the field distributions at the band-edges. Then we created the cavity by changing the minor axis diameter of the holes at the center. We used a linear tapering (Figure 1b), and observed a cavity confined mode. We optimized our design parameters, and in the final design, we have: SiN thickness $t = 330 nm$; width of $w = 450\ nm$; Bragg period $a = 233\ nm$; the major and minor diameter of the elliptical holes are $300\ nm$ and $100\ nm$ respectively. The length of the taper region is $\sim 2.04\ \mu m$ and the major diameter of the ellipses are tapered to $140\ nm$, by keeping the minor diameter same. The gap between two center holes is $115\ nm$. We also tapered down the Bragg periodicity to $222\ nm$ to obtain the best performance. We find that the electromagnetic mode is confined in the cavity region with a mode volume of $\sim 2.5\ (\lambda/n)^3$. This mode-volume is a factor of five larger than floating SiN nanobeam resonators[21, 22] because the mode confinement suffers due to the encapsulation with material of refractive index $\sim 1.5$. However, it is significantly smaller compared to the whispering gallery mode resonators. The highest calculated Q-factor is

~$10^5$ with 40 Bragg mirrors on each side of the taper region, and does not improve further with increased mirrors.

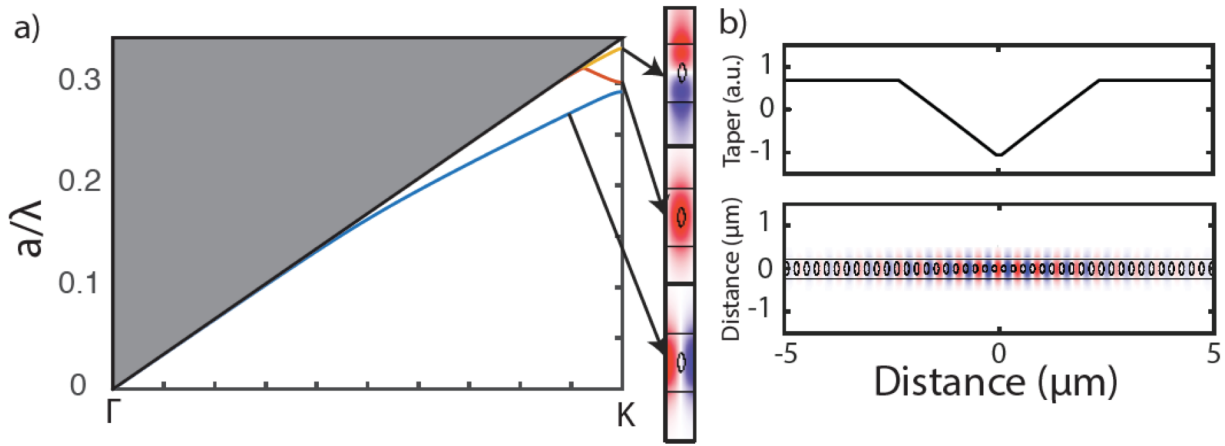

*Figure 1: Design of the encapsulated silicon nitride nanobeam cavity: (a) band diagram of the 1D photonic crystal, where the SiN has a refractive index of 2, and the surrounding medium has a refractive index of 1.47. The field profiles from different bands are also shown. (b) The cavity is realized by linearly decreasing the hole's major radius as well as the hole spacing. An image of the resulting localized field overlayed with the nanobeam edges for clarity.*

**Transmission measurements**

We fabricated the nanobeam cavity using 330 nm thick SiN grown via LPCVD on 4 μm of thermal oxide on silicon. The nanobeam parameters are as stated above, although in our experiments we used 20 Bragg mirrors on both sides so the entire nanobeam could be viewed with a confocal microscope. The increased field of view is critical for our transmission measurement as we pump and collect light from two different gratings in the nanobeam (Figure 2a). We estimate that with such reduced mirrors the simulated Q-factor reduces to Q~15,000. We spun roughly 400 nm of Zeon ZEP520A, which was coated with a thin layer of Pt/Au as a charging layer. The resist was then patterned using a JEOL JBX6300FX with an accelerating voltage of 100 kV. The pattern was transferred to the SiN using a RIE etch in $CHF_3/O_2$ chemistry. Figure 2a shows the SEM of the fabricated SiN cavities just after etching. To encapsulate the nanobeams we spun Poly(methyl methacrylate) (PMMA) and baked the chip at $180^oC$. PMMA and PECVD silicon dioxide have a similar refractive index[23], however PMMA is preferable in our experiments because it can be removed without any risk to the nanobeams. We measured the resonant transmission of the optical resonator using supercontinuum light-source, and clearly identified the band-gap and the cavity mode (Figure 2b). This cavity mode has a Q-factor of ~2,000. We probed the transmission of the nanobeam resonators through the gratings as shown in Figure 2a. When several cavities with linearly scaled periods and major radii were measured, we observed the expected linear scaling of the cavity resonances with periodicity of the nanobeam. The measured cavity Q-factors are in the range 1,500-7,000. We attribute the lower quality factor compared to the theoretical predictions to the fabrication imperfections. We also extended our work to the telecommunication band and fabricated encapsulated nanobeam resonators for 1550 nm (see Supplementary Materials).

**Integration with layered materials**

After the passive characterization, we moved to layered 2D material integration. A monolayer $WSe_2$ was transferred on the SiN nanobeam cavity using the dry transfer method with polycarbonate (PC) stamp[24]. As mentioned earlier, after the transfer of the monolayer, we spun PMMA on top to encapsulate the nanobeam. To avoid the monolayer removal while cleaning PC, we kept PC on the cavity and spun PMMA on top. Figure 3a shows the SEM of the cavity before $WSe_2$ transfer, and Figure 3b shows the SEM of the cavity with 2D material. The SEM in Figure 3b is in false color to highlight the position of the $WSe_2$ monolayer. We measured the transmission of two different nanobeam cavities both with and without 2D material (Figures 3c-f). The cavity modes are shown in the shaded region. Before transferring WSe$_2$, the measured Q-factors for device 1, 2 (Figure 3c, 3e) were 2800 and 1600 respectively. We note that, while several cavities with Q-factor around ~7,000 were measured, they were not in the correct spectral window for integration with 2D materials.

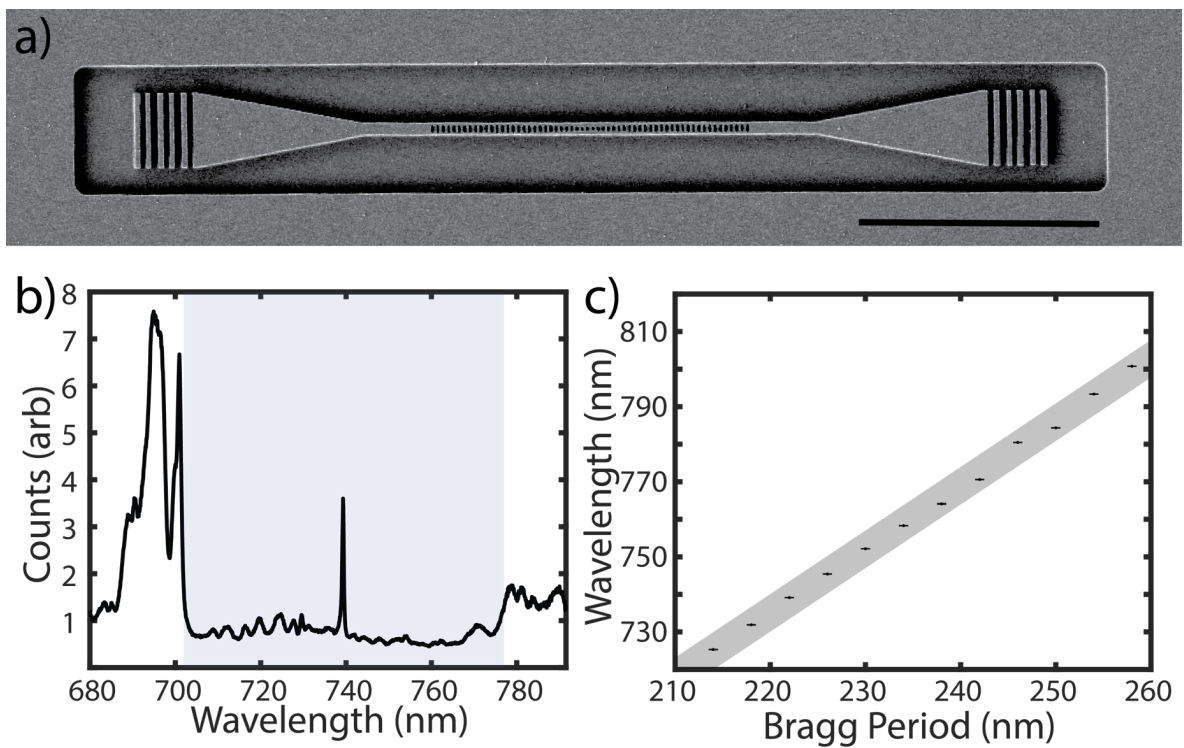

*Figure 2: Bare cavity resonances: (a) An SEM of a fabricated SiN nanobeam prior to encapsulation. The nanobeam resonators is probed via two gratings. The scale-bar is 10μm. (b) Example cavity transmission spectrum as measured through the gratings. The shaded portion highlights the low transmission region from the Bragg reflectors, with the cavity peak at the center. (c) The observed cavity resonances scale linearly with the Bragg period, while holding the radii constant.*

The integration of 2D material leads to red shifts of ~6.7 $nm$ and ~15 $nm$ of cavity modes for devices 1 and 2 as well as reduced Q-factors of 800 and 400, respectively. The linewidth broadening with 2D material on top is expected due to absorption from 2D materials. Additional broadening is also expected because PC and WSe$_2$ can prevent PMMA from filling some of the nanobeam holes. Our numerical simulation shows that the Q-factor of the cavity with few holes without PMMA and PC on top is expected to be ~2,000. This change in the refractive index profile will also affect the resonance wavelength of the nanobeam resonator. However, PMMA not filling the holes would cause a blue-shift in the cavity resonance, rather than the observed red-shift. Therefore, we attribute the red-shift of the cavity modes to the PC (used for transferring 2D materials) on top of the cavity which has a slightly higher refractive index of ($n$~1.57), compared to PMMA ($n$~1.48). We validated our hypothesis by finite difference time domain simulation using Lumerical FDTD Solutions (see Supplementary Materials). We simulated a SiN nanobeam on oxide, and covered by both PMMA and PC. A red-shift with PC on top was observed. In our simulations, we also found that with PC on top, a new cavity mode appears. This is consistent with our experimental findings (Figures 3 d, f). This newly appeared mode is a TM cavity mode (see the mode-profiles in Supplementary Materials), which is attributed to the slightly higher refractive index of PC causing a vertical asymmetry. To further validate that this new mode appears due to PC, we measured the nanobeam transmission with PMMA and PC on top, and without any 2D materials. We observed the appearance of the new mode and red-shift of the cavity (see Supplementary Materials). Note that, a single monolayer has in-plane dipole [25], and thus it interacts primarily with the TE cavity modes. We would expect monolayer WSe$_2$ PL to only couple to the TE mode but not the TM mode.

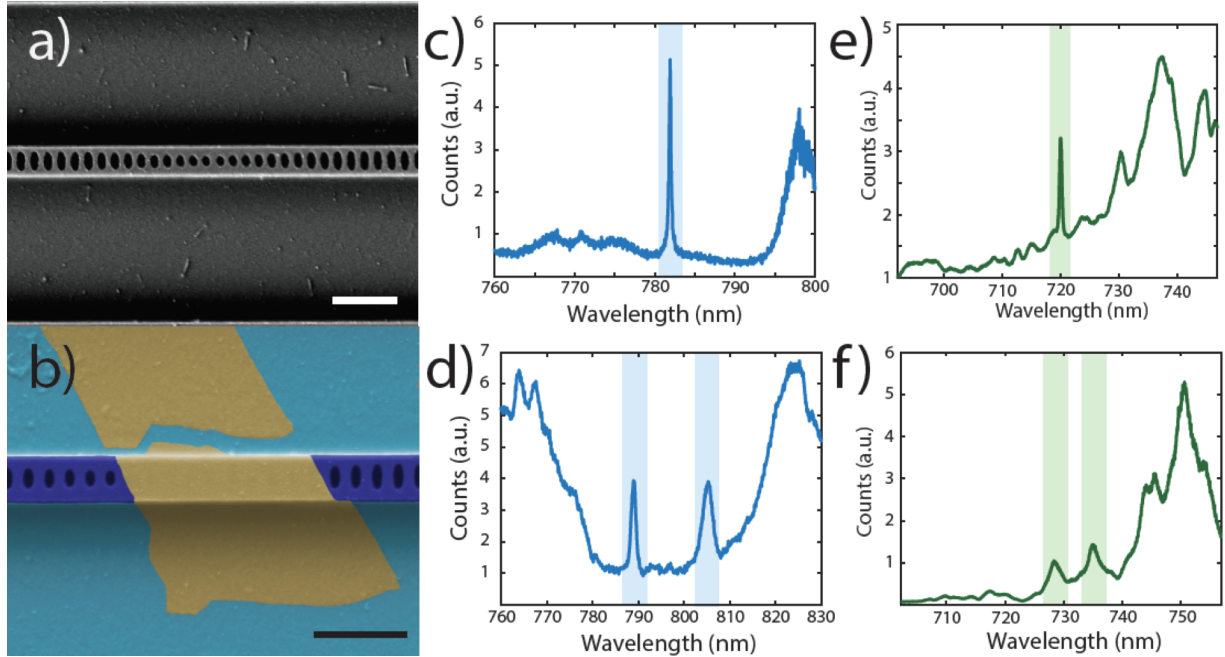

*Figure 3: Transmission through SiN nanobeam before and after transfer of $WSe_2$: (a) A SEM of the defect region of the nanobeam. (b) False colored SEM of a nanobeam with monolayer $WSe_2$. The SiN is shown in dark blue, the silicon oxide is shown in light blue, and the $WSe_2$ is shown in gold. The scale-bar in both figures corresponds to 1µm. (c), (e) The transmission spectrum before transferring $WSe_2$ for devices 1 and 2, respectively. (d), (f) The transmission spectra after $WSe_2$ transfer for devices 1 and 2, respectively.*

Indeed, the measured PL confirms our understanding. We chose cavities with two different wavelengths, to demonstrate coupling with the $WSe_2$ exciton, and with the defects embedded in $WSe_2$ monolayer, which are known to exhibit single emitter like properties[26-29]. Figures 4a and b clearly shows the cavity enhanced exciton and defect state PL, respectively, taken at 80K. The fact of only one peak in the PL spectrum confirms that the monolayer only interacts with the TE mode and the TM nature of the second mode in the transmission spectrum. By fitting a Lorentzian curve to the cavity peaks in the PL spectrum we find Q-factors of 830 and 320 for device 1 and 2 respectively. These linewidths agree with the linewidths observed in transmission.

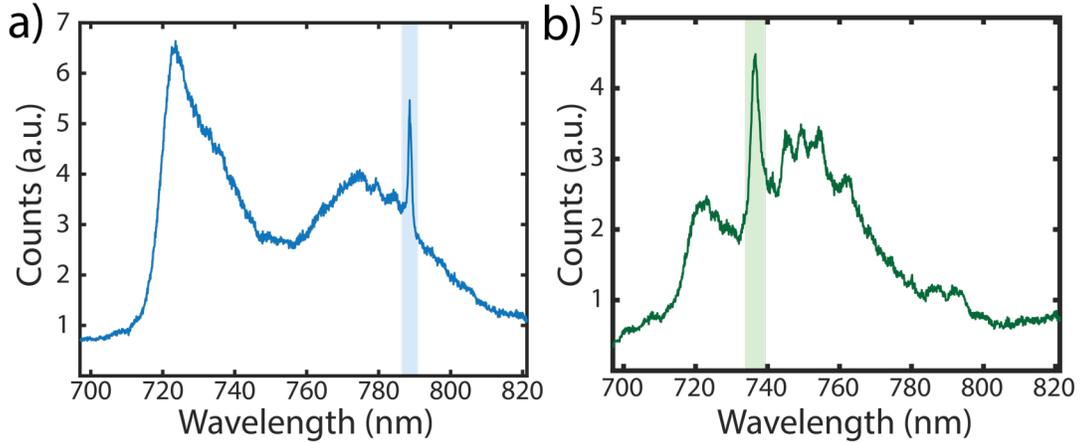

*Figure 4: PL from the WSe$_2$ clad SiN nanobeam resonators: (a) PL from device 1 shows cavity enhanced PL from the defects in WSe$_2$ monolayer and (b) PL from device 2 shows cavity enhanced PL from the WSe$_2$ exciton.*

**Discussion**

We demonstrated the operation of an encapsulated SiN nanobeam despite the low index contrast between the SiN and the material it is encapsulated within. Encapsulation provides the desired mechanical stability, which is crucial to build a hybrid optoelectronic platform with emerging materials integrated on passive integrated photonic circuits. Such stability is absent in conventional floating membranes of SiN nanobeam resonators. Our analysis shows that such encapsulation results in a five-fold enhancement of the mode-volume, but the Q-factor of the cavity can still reach $\sim 10^5$. To demonstrate the efficacy of the developed resonators for 2D material integration, we transferred monolayers of WSe$_2$ with these resonators and observed enhanced PL from both the 2D exciton and the defects embedded in the 2D materials. Going beyond layered 2D materials, we envision the encapsulated SiN nanobeam can be used to enhance the light-matter interaction with other emerging nano-materials, such as solution processed emitters and chromophores[30], and Perovskite light sources[31]. We note that, the encapsulation naturally provides a way to preserve the quality of materials, which are sensitive to environmental conditions. By virtue of their small mode-volumes, encapsulated SiN nanobeam resonators will provide a useful tool to develop hybrid large-scale photonic integrated circuits, with applications in optical information science and sensing.

AUTHOR INFORMATION

**Corresponding Author**

* To whom correspondence should be addressed.  Email:arka@uw.edu

**Author Contributions**

T.K.F. conceived the idea. Y.C. performed the simulation. T.K.F. fabricated the SiN resonator. T.K.F. and Y.C. fabricated the 2D material-cavity device. J.W. and Z.M.P. helped with finding the 2D materials and transferring them. T.K.F. and Y.C. performed the optical characterization. T.K.F. wrote the paper with input from everyone. X.X. and A.M. supervised the whole project.

**Funding Sources**

This work is supported by the National Science Foundation under grant NSF-EFRI-1433496, and the Air Force Office of Scientific Research-Young Investigator Program under grant FA9550-15-1-0150. XX acknowledges support from AFOSR (FA9550-14-1-0277). All the fabrication processes were performed at the Washington Nanofabrication Facility (WNF), a National Nanotechnology Infrastructure Network (NNIN) site at the University of Washington, which is supported in part by the National Science Foundation (awards 0335765 and 1337840), the Washington Research Foundation, the M. J. Murdock Charitable Trust, GCE Market, Class One Technologies and Google.

**ACKNOWLEDGMENT**

We thank Mr. Richard Bojko, and Mr. Alan Zhan for helpful discussion about SiN fabrication.

**Supporting Information Available**: This material is available free of charge via the Internet at http://pubs.acs.org.

# Supplementary material: Encapsulated silicon nitride nanobeam cavity for nanophotonics using layered materials

Taylor K. Fryett[1], Yueyang Chen[1], James Whitehead[1], Zane Matthew Peycke[2], Xiaodong Xu[2,3], Arka Majumdar[1,2]

[1] Electrical Engineering, University of Washington, Seattle, WA 98189, USA

[2] Department of Physics, University of Washington, Seattle, WA 98189, USA

[3] Materials Science and Engineering, University of Washington, Seattle, WA 98189, USA

## S1. Broken floating membranes:

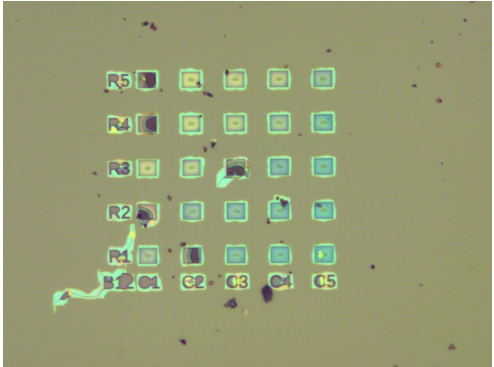

*Figure S1: Broken floating GaP photonic crystal cavities.*

Many photonic crystal cavities are thin floating membranes. These membranes are susceptible to the mechanical stress associated with many processes. For instance, several of the gallium phosphide photonic crystal cavities in Figure S1 are visibly broken. In particular, cavity R2C1's lower left corner has broken away from the main membrane, and the curvature of the cavity can be seen in the interference fringes on the main body of the cavity. Most of these cavities are broken during 2D material transfer process.

## S2. Encapsulated SiN nanobeam resonators at the telecommunications band:

Another important wavelength for 2D materials and photonics is the telecommunication band (near 1550nm). By scaling our design for 750 nm to telecommunication wavelengths we found a suitable design for SiN encapsulated nanobeam resonator. Our devices show excellent rejection of light within the band-gap of the photonic crystal with the exception of a lone cavity peak (Figure S2.a). By fitting a Lorentzian function to the cavity peak we find that a typical device has a quality factor around ~10,000.

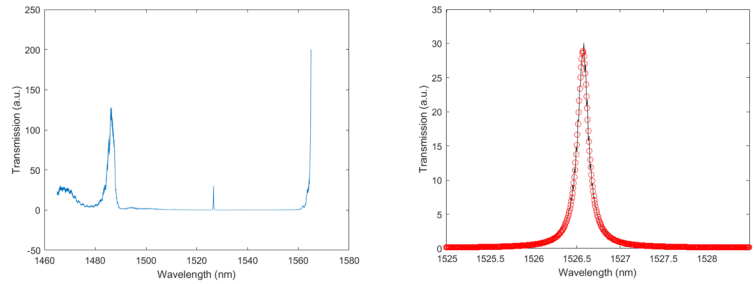

*Figure S2: Cavity Transmission at Telecommunication wavelength. (a) Broadband cavity transmission showing the isolated cavity peak within the photonic bandgap. (b) Closeup of the cavity resonance with a Lorentzian fit.*

## S3. Effect of PC and PMMA as encapsulation layer

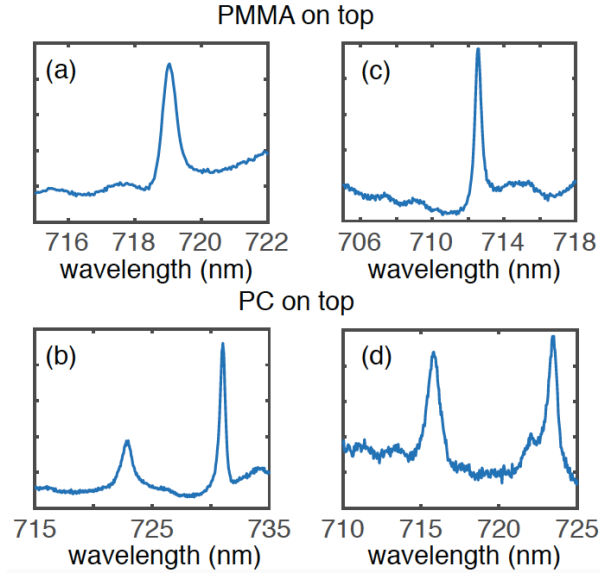

***Figure S3:*** *Measured transmission spectrum of the SiN nanobeam cavities with PMMA and PC on top. We observe a red-shift with PC, and a new mode appears.*

As we explained in the main text, after the 2D material transfer we measured an additional mode in transmission, although, the mode is visible in the PL measurement. To understand the origin of this mode, we measured cavity modes with PMMA and PC spun on the same structures. We measured several devices, and results from two characteristic devices are shown in Figure S3. Note that, these are two different cavities from the ones already reported in the main text.

To understand the origin of the new mode, we performed FDTD simulation using commercially available Lumerical FDTD solutions. We find that due to the slightly higher index of PC compared to PMMA and oxide, the PC-clad nanobeam has a vertical asymmetry. This leads to appearance of a new TM mode. We did observe two different confined modes with PC on top. Figure S4 shows the distribution of the field of the confined mode with PMMA on top (the out-of-plane direction is the z-direction). Figure S5 and Figure S6 show the field distributions for two cavity modes with PC-clad nanobeam resonators, clearly showing one of the mode is TM with strong field along *z*-direction. This mode, however, will not couple well with the 2D material placed on top as the monolayer primarily has in-plane mode.

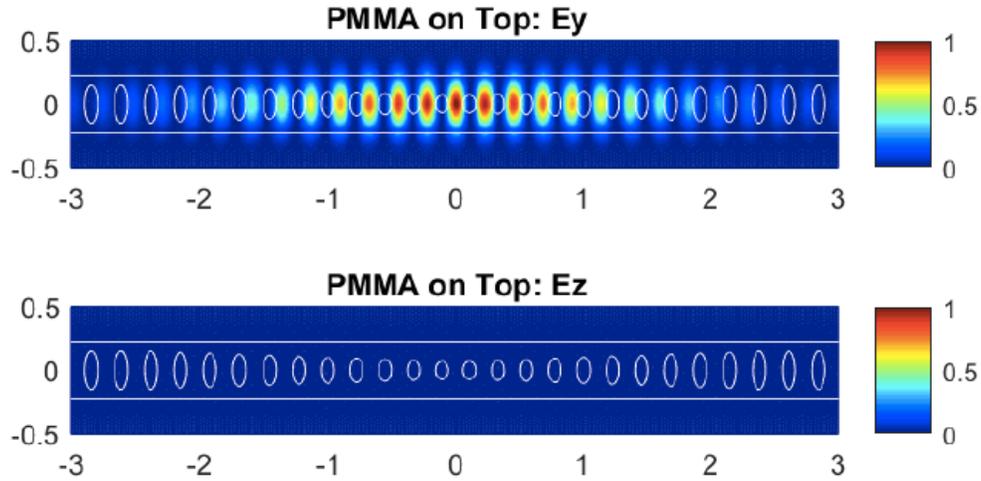

*Figure S4:* Field distribution of the PMMA-clad nanobeam cavity. The field along z-direction is almost zero, and the confined field is along the y-direction.

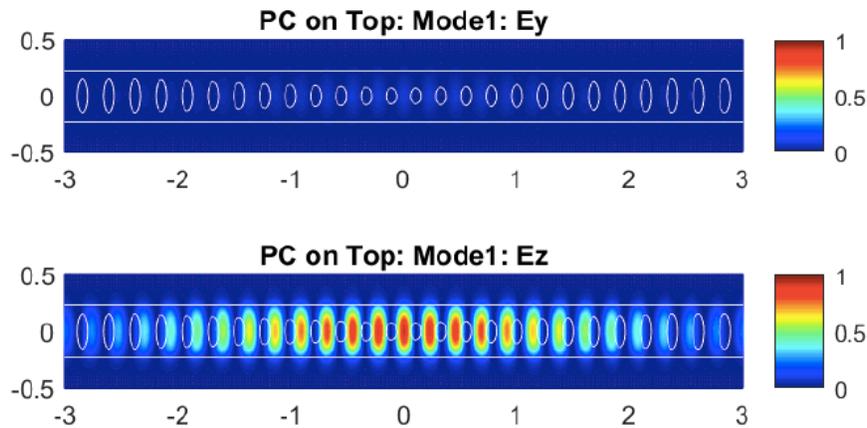

*Figure S5:* Field distribution of the PC-clad nanobeam TM cavity. This is the new TM modes, with large field along z-direction.

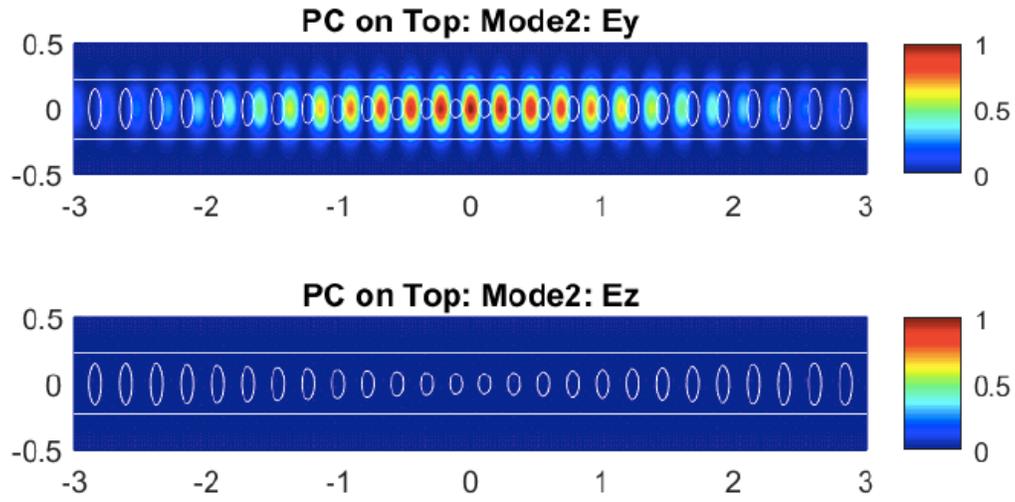

*Figure S6:* *Field distribution of the PC-clad nanobeam TE cavity. The field along z-direction is minimal.*